# Deep Learning Applications for Lung Cancer Diagnosis: A systematic review

Hesamoddin Hosseini [a], Reza Monsefi [a], Shabnam Shadroo [b,*]

[a] Department of Computer Engineering, Ferdowsi University of Mashhad, Mashhad, Iran
[b] Department of Software Engineering, Islamic Azad University, Mashhad Branch, Mashhad, Iran

Email: hesamoddin.hosseini@mail.um.ac.ir, monsefi@um.ac.ir, sh_shadroo@mshdiau.ac.ir

* Corresponding Author

## Abstract
Lung cancer has been one of the most prevalent disease in recent years. According to the research of this field, more than 200,000 cases are identified each year in the US. Uncontrolled multiplication and growth of the lung cells result in malignant tumour formation. Recently, deep learning algorithms, especially Convolutional Neural Networks (CNN), have become a superior way to automatically diagnose disease. The purpose of this article is to review different models that lead to different accuracy and sensitivity in the diagnosis of early-stage lung cancer and to help physicians and researchers in this field. The main purpose of this work is to identify the challenges that exist in lung cancer based on deep learning. The survey is systematically written that combines regular mapping and literature review to review 32 conference and journal articles in the field from 2016 to 2021. After analysing and reviewing the articles, the questions raised in the articles are being answered. This research is superior to other review articles in this field due to the complete review of relevant articles and systematic write up.

## Introduction
The most dangerous type of cancer is lung cancer which has had the highest mortality rate. Early detection can save many lives [1]. Lung cancer is the second most common cancer, with prostate cancer in men and breast cancer in women.

The International Association of Cancer Society (IACS) estimates the number of lung cancers in the United States for 2020 as follows [2]:
- About 235,760 new cases of lung cancer (119,100 in men and 116,660 in women)
- About 131,880 deaths from lung cancer (69,410 in men and 62,470 in women)

Pulmonary glands are small lesions inside the lungs that are usually round and can cause lung cancer if diagnosed late. Lung cancer in the early stages is asymptomatic on CT scan due to its small size and location of the glands, and its symptoms appear when the disease is in more advanced stages.

Computed Tomography (CT) and Magnetic Resonance Imaging (MRI) are standard medical procedures for early detection that improve patient survival [3], [4]. Previous intelligent methods used hand-drawn feature extraction methods, such as Sequential Flood Feature Selection Algorithms (SFFSA) or Genetic Algorithm (GA), which can help generate optimal features [5]. In recent years, CAD systems have used deep learning technology that automatically extracts image features [6], and many medical image processing programs have been successful due to the use of deep learning technology [7].

The two main types of lung cancers are small-Cell Lung Cancer (SCLC) and Non-Small Cell Lung Cancer (NSCLC). Things that lead to lung cancer: Smoking - in both smokers and people exposed to second hand smoke, toxic particles in the air, sex, genes, aging, etc [8]. The main cause of lung cancer is prolonged smoking as shown in the figure. 1 [9].

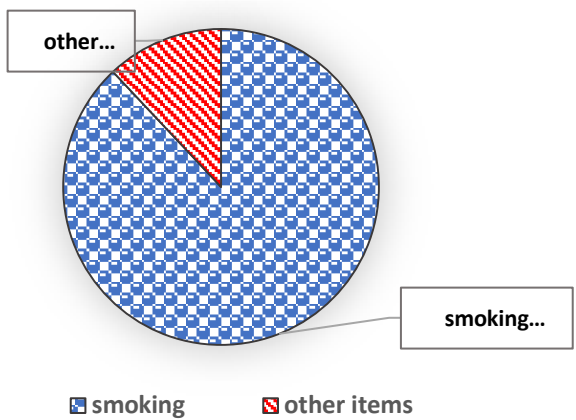

Fig. 1. The Main Cause of Lung Cancer

Symptoms used to diagnose lung cancer including, yellow fingers, anxiety, chronic illness, fatigueness, allergies, wheezing, roaring, coughing up blood, even small amounts, hoarseness, shortness of breath, bone pain, headache, difficulty in swallowing and chest pain [10].

Various models have been proposed for the diagnosis of early-stage lung cancer, such as "Improved Profuse Clustering Technique Deep Learning with Instantaneously Trained Neural Networks (IPCT-DLITNN)" [11], "Adaptive Hierarchical Heuristic Mathematical Model (AHHMM)" [12], "Artificial Neural Network (ANN)" [13], "Deep Convolutional Neural Network (DCNN)" [14], etc. which are reviewed in this article. Each different model leads to a different level of accuracy and sensitivity.

In the proposed taxonomy, there are different models that lead to different accuracy and sensitivity for diagnosing lung cancer. In this research, we categorize the presented articles into 5 sections. Diagnosis of lung cancer through deep learning consists of five parts. These are the types of inputs, pre-processing methods, architecture, hybrid method and transfer learning shown in Figure 2.

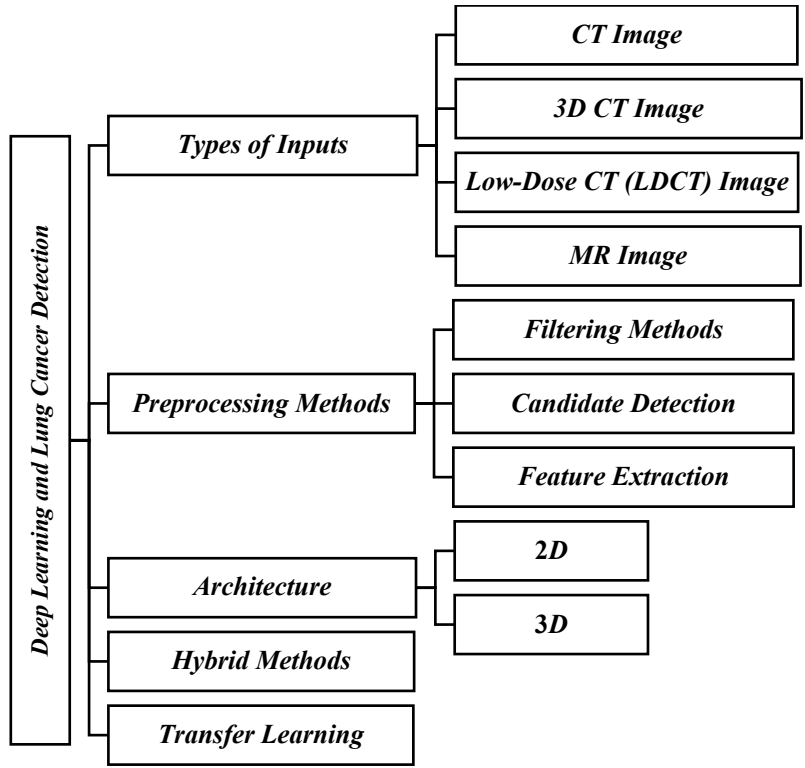

Fig. 2. The Taxonomy of Lung Cancer Detection

There is a gap in research that should systematically review existing articles on lung cancer based on deep learning approach, as it has not been systematically reviewed before. We will ask a few questions at the beginning of this article. At the end, according to the created analysis, the answers to the questions will be provided. By reviewing the articles presented, we have reviewed the countries and universities active in this field.

The rest of the article is organized as follows. Section 2 describes the proposed research method and research questions. Section 3 main studies and Section 4 analysis based on literature review. In addition, Section 5 answers research questions. Finally, Section 6 presents the conclusion.

## 2. Research Methodology
In this paper, the methods are examined with a systematic approach. The steps used are similar to the steps used in the Cruz-Benito article [15] and some other articles [16,17, 18]. Steps such as Systematic Mapping Study, (SMS) and Systematic Literature Review, (SLR) are mentioned in this article. The steps performed in the study will be further explained.

## 2-1: RQs (Research Questions)
Systematic review is a term that has specific meaning and refers to a specific process in academic writing. This is the standard for quality and legal academic review articles. This type of article helps readers come up with creative ideas.

Research Questions (RQ) are given follows:

**RQ1:** How Many articles on SMS and SLR have been published on lung cancer from 2016 to 2021?
The answer to this question helps researchers understand the history of lung cancer.

**RQ2:** What percentage of the articles on lung cancer diagnosis are based on deep learning from 2016 to 2021?
The answer to this question introduces researchers to other areas of lung cancer and provides a percentage.

**RQ3:** What are the best tools and simulations for deep learning-based lung cancer?
The answer to this question helps researchers understand the tools used in this field.

**RQ4:** What is the best dataset for lung cancer based on deep learning?
The answer to this question helps researchers understand the data sets used in this field.

**RQ5:** What universities have worked on the diagnosis of lung cancer based on deep learning?
The answer to this question helps researchers to get acquainted with the universities that worked in this field.

**RQ6:** What kind of deep learning algorithms are used to classify an image?
The answer to this question helps researchers to have different algorithms for classification.

**RQ7:** According to the question 6, what is the accuracy or sensitivity of this algorithm?
The answer to this question helps researchers become familiar with the different percentages of accuracy and sensitivity of algorithms.

## 2-2: Data Base

This article has been searched in the following scientific databases and books have been cited in them.

Databases are used in articles, we have selected and used some of them [15]. The electronic databases used in this research are:

- *IEEE Xplore* (*http://www.ieee.org/web/publications/xplore/*)
- *ScienceDirect – Elsevier* (*http://www.elsevier.com*)
- *Springer Link* (*http://www.springerlink.com*)
- *Wiley Inter Science* (*https://onlinelibrary.wiley.com*)
- *Google Scholar* (*https://scholar.google.com*)

## 2-3. Search Terms

Our first study on this topic was a review article on this topic because it gave us an overview of the subject. The terms used in Table 1 were used to search for the obtained article, and this phrase was the same in all databases. Then we went to search for the subject mentioned in the databases. Finally, the abstracts of the articles were studied and labelled as a survey article. We left out articles that did not have topics such as survey, challenges, reviews and literature, SMS and SLR, and only introduced and reviewed the methods.

Table 1. Searched Terms

| | |
|---|---|
| S1 | (("Document Title":*Lung Cancer*) AND "Abstract":*Deep Learning*) |
| S2 | (("Document Title":*Lung Nodule*) AND "Abstract":*Deep Learning*) |

## 2-4: ICS (Inclusion Criteria Search) and ECS (Exclusion Criteria Search)

Inclusion criteria and exclusion criteria aim to identify studies related to more details. We reviewed all English articles in journals and conferences from 2016 to 2021. Table 2 of the articles explains the Inclusion Criteria and Exclusion Criteria papers.

Table 2. ICS and ECS

| **Inclusion Criteria** |
|---|
| *Studies on different methods of early-stage lung cancer* |
| *Studies have been published from 2016 to 2021* |
| **Exclusion Criteria** |
| *Studies that focus on books and technical reports* |
| *Studies that are not in English* |
| *Studies that are not related to research questions* |
| *Studies in which claims focus rather than evidence* |

## 2-5. Review Phases

The process of choosing articles was carried out through these steps:
1- Articles are looked for using the search terms especially in different databases.
2- A number of them are pruned out by the Exclusion Criteria.
3- After reading the titles and abstracts of the articles the irrelevant ones are subtracted.
4- Finally, after studying the articles in its entirety the main articles are prepared.

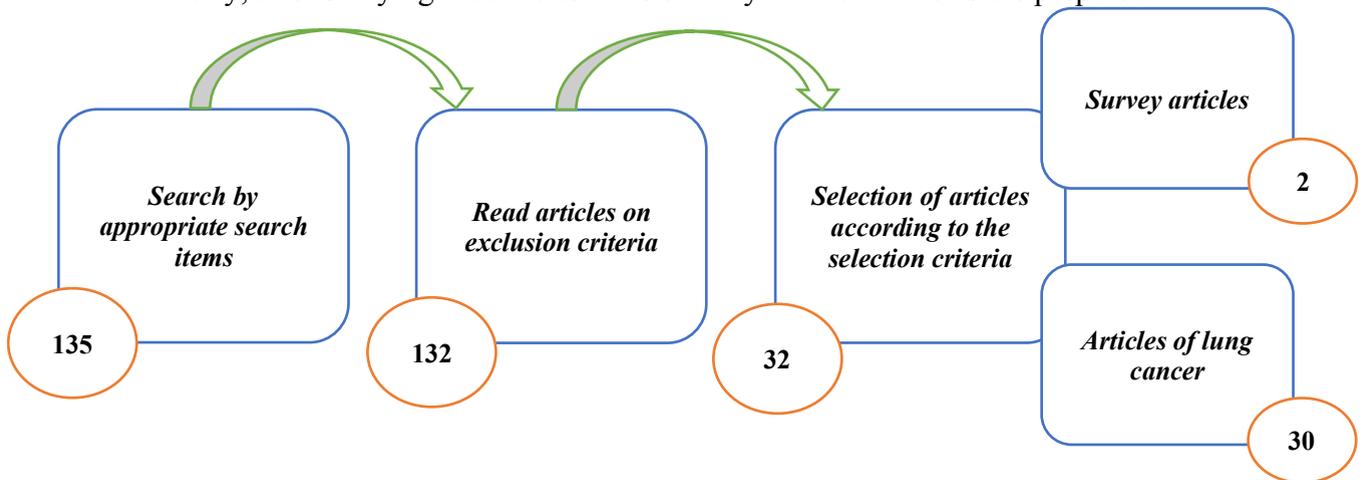

Fig. 3. Article Selection Processes in the Study

From 2016 to 2021, there are 135 articles on lung cancer in the listed databases. Figure 4 shows the number of articles and Figure 5 shows the number of articles in journal and conference per year.

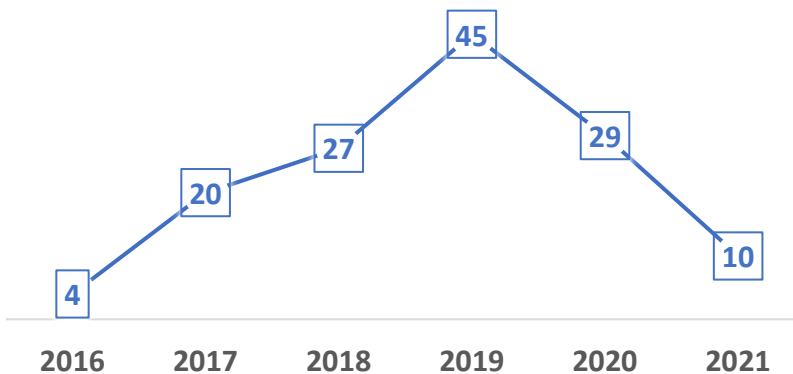

Fig. 4. The Number of Articles on Lung Cancer by Year

As shown in Figure 4, in 2019 the highest study and in 2016 the lowest study in this field. This shows that this issue is increasing every year, which we think is of particular importance.

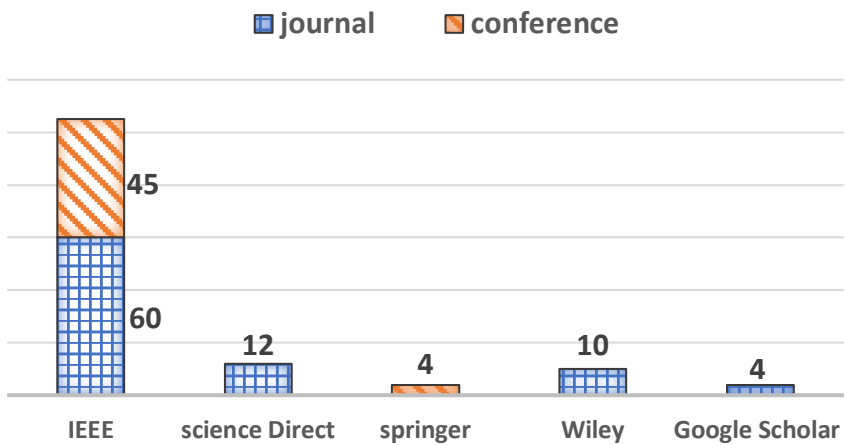

Fig. 5. The Number of Journal and Conference Articles on the Lung Cancer

As can be seen in Figure 5, the IEEE has the highest published papers on lung cancer based on deep learning. Springer and Google Scholar also have the fewest published articles in this field.
After studying the abstracts and titles of 135 articles in the field of lung cancer, 32 articles were selected and fully studied. Figure 6 shows the number of selected articles on lung cancer by year and Figure 7 shows the number of journal and conference articles in this field.

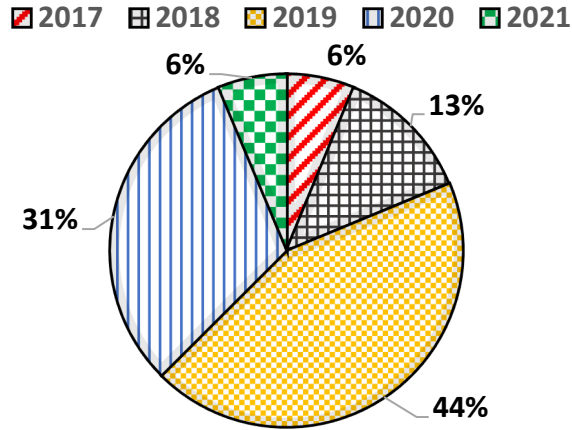

Fig. 6. The Percentage of Selected Articles on the Lung Cancer

As shown in Figure 6, the highest percentage of our study on lung cancer based on deep learning is in 2019 and the lowest percentage is in 2017.

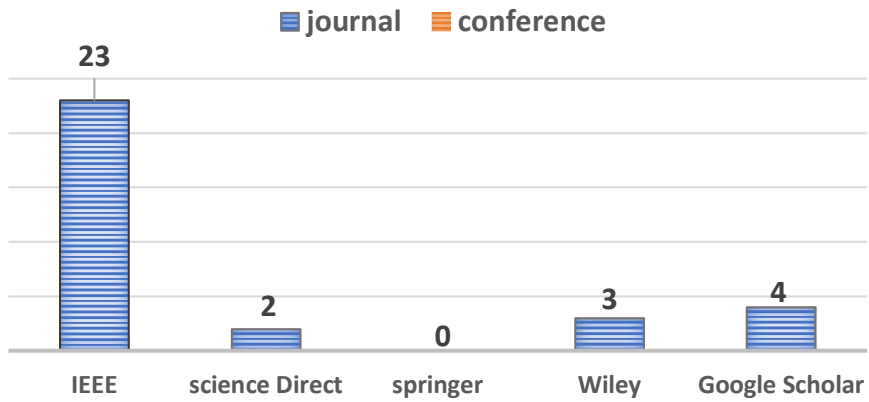

Fig. 7. The Number of Selected Journal and Conference Articles on the Lung Cancer

As shown in Figure 7, most of our deep learning of lung cancer is in the IEEE. Springer also has the lowest number of our studies in this area.

## 3. Literature review

In 2021, methods that use gene expression data are expensive but highly accurate. In contrast, there is a radiometric method that is cost-effective though its accuracy is not competitive. P. Aonpong et al. [19] suggested a genotype-guided radiomics method (GGR) that results high accuracy and low cost. This method is performed sequentially: pre-processing, radiomics feature extraction and selection (input features), prediction. This prediction, called GGR, consists of two steps that use two models. The first model uses gene estimation and the second model predicts recurrence using the estimated gene. This method uses the general NSCLC radiographic data set, which includes CT images and gene expression data. Experimental results of this method show that the prediction accuracy can be significantly improved from existing radiometric method and ResNet50 to 83.28% by the proposed GGR.

Statistics show that the main cause of disease and death in lung cancer is how it is diagnosed in the early stages. To evaluate the mutagenic status of Epidermal Growth Factor Receptor (EGFR), F. Silva et al. [20] recommended MLP for the final classification of EGFR mutation status. This EGFR assessment includes the nodule, the lung containing the main nodule, and both lungs. This proposed approach consists of two main phases. The first phase is the feature learning task. The second phase is an end-to-end classification model based on transfer learning techniques. This method uses the LIDC-IDRI and NSCLC-Radiogenomics datasets. Experimental results show that it has the best ability to predict.

In 2020, the leading cause of lung cancer death and disease is how it is diagnosed in the early stages. H. Yu et al. [12] proposed the "Adaptive Hierarchical Heuristic Mathematical Model (AHHMM) " for the automatic diagnosis of lung cancer. This method consists of five steps. The first step is to get the image. The second stage is Pre-processing. The third step is Binarization. Next, Thresholding and segmentation. Finally, feature extraction and detection by a Deep Neural Network (DNN). Modified K-means clustering was also used for pre-classification images. Experimental results show that this method has an accuracy of 96.67% of the lung cancer dataset [21].

Radiologists often use screening for a wide range of CT scans for accurate examination. Automated algorithmic solutions may help, but the relationship between algorithmic solutions and physicians is also a challenge. To solve this problem, a system called low-dose CT scans has been proposed by O. Ozdemir et al. [22] This system directly analyzes the CT scans and provides calibrated scores. These are system-based three-dimensional convolutional neural networks. This method is performed sequentially: Pre-process, Computer-Aided Detection module (CADe) for segmentation, Computer-Aided Diagnosis module (CADx). CADx depends on the performance of the CADe, so it is developed and configured simultaneously. This method uses LIDC-IDRI, LUNA-16, Kaggle datasets. This system is more reliable in the real world. The proposed model can diagnose lung cancer with 96.5% accuracy.

Q. Zhang et al. [23] stated that many can be saved if lung cancer is diagnosed in the early stages. Early detection of lung cancer nodules is a very challenging, time-consuming, and repetitive task for radiologists. They proposed a system called a "Multi-Scene Deep Learning Framework (MSDLF)" with the "vesselness filter" to increase the accuracy and decrease the false positive criteria. The main purpose of this analysis is to determine large nodes (> 3 mm). A model designed by four-channel CNN. The method includes the following steps: Preparation of data set, mending of lung contour and segmentation of parenchyma, the removal of the vessel, the data set standardization, design of the CNN Design, segmentation, and classification, normalized spherical sampling. The LIDC-IDRI dataset is used in this method.

A. Masood et al. [24] stated that drawing lung nodules manually by radiologists is a time-consuming and tedious task. To help radiologists, the system a 3D Deep Convolutional Neural Network (3DDCNN) is provided. Their system works better than advanced systems. The combination of deep learning and cloud computing is used to accurately detect lung nodules. They used the Multi-Region Proposal Network (mRPN) in their architecture. The method consists of training Datasets, data augmentation, pre-processing, proposed model architecture, training process, cloud-based 3DDCNN CAD system. The ANODE09, LUNA-16, LIDC-IDRI,

SHANGHAI Hospital datasets are used in this method. The evaluation of the experiments shows that the presented model can detect 98.5% accuracy of lung cancer.

H. Guo et al. [25] proposed the Knowledge-based Analysis of Mortality Prediction Network (KAMP-Net) to predict the risk of mortality in lung cancer. In this method, data augmentation is used to train Convolutional Neural Network (CNN). They supposed that the data augmentation is effective for improving the performance of CNN. The clinical measurements are used to train a Support Vector Machine (SVM) classifier and the CNN and SVM results are combined to generate a risk of mortality. The clinical measurements have been manually acquired. The method includes the following steps: Multi-Channel Images Coding, Network Design, and Implementation, Integration of Deep Learning, and Clinical Knowledge. The National Lung Screening Trial (NLST) dataset is used in this method.

S. Pang et al. [26] used deep learning to identify the type of lung cancer in CT images of patients at Shandong Province Hospital. For solving the low data problem collected by the patient, they used image preprocessing methods such as "rotation, translation, and transformation" in order to extend training data. The authors trained the "densely connected convolutional networks (DenseNet)" to categorize the lung cancer images.
Finally, they use the adaptive boosting (adaboost) algorithm for aggregating multiple classification results. The Shandong Provincial Hospital dataset is used in this method. The evaluation of the experiments shows that the proposed model can detect 89.85% accuracy of lung cancer.

Automated lung node analysis needs both an accurate classification of malignant lung nodes and feature score regression. L. Liu et al. [27] proposed the MTMR-Net model for the automatic analysis of pulmonary nodules. In addition to this, the architecture of the Siamese network is designed in this model. The proposed method of architecture is comprised of three main modules. The first module is feature extraction module. It comprised of a convolution layer, a Res Block A, and three Res Block B. The second module is classification module. It comprised of only one fully connected layer. The third module is regression module. It comprised of two fully connected layers. The MTMR-Net model includes Multi-Task Learning for Lung Nodule Analysis, Margin Ranking Loss for Discriminating Marginal Nodules, Joint Training of MTMR-Net. The evaluation of the experiments shows that this proposed model can detect 93.5% accuracy of the lung cancer dataset [28]. Comparing with other advanced methods, "MTMR-NET" achieved the best accuracy, sensitivity, specificity.

J. Zheng et al. [29] stated that the main cause of disease and death in lung cancer is how it is diagnosed in the early stages. Risk classification of pulmonary nodules in adenocarcinoma is necessary for the early detection of lung cancer. To improve the diagnosis of pulmonary adenocarcinoma in this article. They presented STM-Net that a deep Convolutional Neural Network (CNN) with a scale transfer module (STM). The method includes the following steps. Firstly, images of the input pulmonary nodule pass through four layers of convolution. Secondly, using max-pooling and STM to unify the size of feature maps. Thirdly, use channel fusion to achieve the final classification. The Zhongshan Hospital Fudan University (ZSDB) dataset is used in this method. The network was trained and tested to predict the hazard phase. The experimental results show that this method has an accuracy of 95.455% on successfully detecting lung cancer.

L. Cai et al. [30] presented a 3D visualization method using "Mask Region-Convolutional Neural Network (Mask R-CNN) and a volume rendering algorithm that ray-casting" for detection and segmentation for the pulmonary nodule. This method is comprised of three modules. The first module is the Preprocessing Module (PrM). The second module is the Segmentation Module (DSM). Finally, a three-dimensional Reconstruction Module (3DRM). Mask R-CNN mainly is comprised of four parts. The first part is resnet50 as the backbone was a multilayer neural network. The second part is the feature pyramid network (FPN). The third part is Region Proposal Network (RPN). The fourth part is Function branches. They used resnet50 to extract the feature maps of the original image and FPN to fully use multi-scale feature maps and used RPN to get candidate bounding boxes. Then, the candidate box was applied to normalize RoIs by Region of Interest (RoI) Align. The LUNA-16, Ali TianChi challenge datasets are used in this method. The evaluation of the experiments shows that this proposed model can detect 88.7% sensitivities of lung cancer.

In 2019, The exact identification of benign-malignant lung nodules is very important in the early diagnosis of lung cancer. Y. Xie et al. [31] recommend a deep neural network model to separate benign-malignant lung nodules in a multi-view knowledge-based collaborative (MV-KBC). First, they break each 3D lung node into nine fixed views to learn the properties of the 3D node. Next, for each view, they create a knowledge-based collaborative (KBC) sub-model, where the overall appearance (OA), heterogeneity in voxel values (HVV), and heterogeneity in shapes (HS) patches extraction are designed for lung nodules on each slice, respectively. Finally, nine KBC sub-models are commonly used to classify nodes with a consistent weighting scheme. Thus, the "MV-KBC model" trains them with an end-to-end approach. The LIDC-IDRI dataset is applied in this method. The proposed model can detect 91.60% accuracy of lung cancer.

Y. Li et al. [32] proposed a method for thoracic MR images based on deep learning. Because MR images were performed with few methods of detecting. Unlike many nodule detection methods for CT images, the proposed method takes the whole image as input and didn't require candidate extraction. As the lung nodules are very different in size. Faster R-CNN is designed for lung nodules detection to prevent candidate extraction and less dependence on a scale. The Faster R-CNN is comprised of two modules. The first module can generate proposed regions for each image, named a Region Proposal Network (RPN). The second module can classify the region proposals, named a Fast R-CNN detector. The First Affiliated Hospital of Guangzhou Medical University dataset is used in this method. The evaluation of the experiments shows that this proposed model can detect 85.2% sensitivities of lung cancer.

Early and accurate diagnosis of Small cell lung cancer (SCLC) is critical to improving survival. Accurate cancer segmentation helps doctors make better diagnostic decisions. However, manually segmenting is a time-consuming and challenging task. W. Chen et al. [33] proposed a "hybrid segmentation network (HSN)" based on a "convolutional neural network (CNN)". The proposed method of the network is designed of two mains essential for accurate cancer segmentation. The first is a lightweight 3D CNN to learn long-range 3D contextual information. The second is a 2D CNN to capture fine-grained semantic information. They proposed a multiscale separable convolution (MSC) Block is similar to architecture with S3D convolution and a "hybrid features fusion module" (HFFM) for merging the 2D and 3D features. The Hospital Affiliated to Shandong University dataset is used in this method. The experimental results show that this method has a mean accuracy of 0.909 and a mean sensitivity score of 0.872.

The main indicators of malignancy in the diagnosis of lung cancer are the size and shape of the nodule. However, effectively obtaining node structural information from CT scans is a challenging task. P. Sahu et al. [34] proposed a multiple view sampling-based multi-section CNN architecture. The Multi-section CNN is comprised of two features. The first feature is didn't require tedious manual spatial annotation. The second feature is lightweight that this is possible can be easily ported to Mobile devices. The Multi-section CNN based on spherical sampling is comprised of five steps. The first step is the cross sections nodule. The second step is shared parameters across all sections. The third step is the elementwise maximum pooling of features from all sections. Next, is the retraining of the final layer. Finally, the nodule classification. The LIDC-IDRI dataset is used in this method. The proposed model can detect 93.18% accuracy of lung cancer.

Accurate and automatic segmentation of lung nodes holds significant importance in diagnosing lung cancer with aid of computer. W. Wang et al. [35] presented a new "region-based network (Nodule-plus R-CNN)" for detection of pulmonary nodules visible in 3D CT images, which is created with a pile of convolutional blocks. In order to reduce 3D neural network training and annotation efficiently, they presented a new Deep Self-paced Active Learning (DSAL) strategy. The DSAL strategy is combined with "Active Learning (AL)" and "Self-Paced Learning (SPL)" methods. The LIDC-IDRI dataset is used in this method. The evaluation of the experiments shows that this presented model attains 0.66 Dice and 0.96 TP Dice.

The main cause of mortality in lung cancer is how it is diagnosed in the early stages, which use deep reinforcement learning is effective for lung cancer detection. Z. Liu et al. [36] proposed several deep reinforcement learning models that the main idea in these models are to use a multi-layer neural network. The proposed deep reinforcement learning is comprised of two models. The first model is Deep Q-Network (DQN). It comprised of a convolutional neural network with three convolutional layers and two fully-connected layers and the Q-learning algorithm. The second model is Hierarchical Deep Q-Networks (H-DQN). It comprised of two modules. The first module is a top-level module (meta-controller). The second module is a bottom-level module (controller).

A Computed tomography (CT) is used to identify and locate the tumor of cancer. Lakshmanaprabu S.K et al. [37] presented method for classification Computed Tomography (CT) images of lung, named Optimal Deep Neural Network (ODNN). The presented ODNN shows a better classification of lung CT images compared to other classification methods. The classification of Computed Tomography (CT) Images has a different step such as preprocessing, feature extraction, reduction, classification. The ODNN is comprised of two features. The first feature is the manual label time reduction. The second feature is preventing human error. Also, they are used Modified Gravitational Search Algorithm (MGSA) for the optimized structure. The experimental results show that this method has an accuracy of 94.56% and a sensitivity of 96.2% of the lung cancer dataset [38].

Recently, deep learning algorithms have been considered a promising method in the medical field. To identify and classify pulmonary nodules from CT images, C. Zhang et al. [39] designed a three-dimensional convolutional neural network (CNN).
The proposed method is comprised of three modules. The first module is the preprocessing module. The second module is the segmentation and 3D lung reconstruction. The third module is the image enhancement. The LUNA-16, Kaggle, Guangdong Provincial People's Hospital datasets are used in this method.  The experimental results show that the performance of a three-dimensional convolutional neural network (CNN) model has a sensitivity of 84.4% which on superior to manual assessment.

To determine if lung cancer cells are absent or present in human body, I. M. Nasser et al. [13] developed an Artificial Neural Network (ANN) that is similar to a neural network and presentation a quite good technique. An ANN of architecture is comprised of fourth main layers. The first layer is the input layer. The second layer is the one hidden layer. The third layer is the two hidden layers. The fourth layer is the output layer. The experimental results show that this method has an accuracy of 96.67% of the lung cancer dataset, named survey lung cancer [40]. This assessment showed that the neural network is able to diagnose lung cancer.

P. M. Shakeel et al. [11] proposed the improved profuse clustering technique (IPCT) and Deep Learning with Instantaneously Trained Neural Networks (DITNN) approach to improve lung image quality and diagnose lung cancer. The IPCT is comprised of two features. The first feature is Image noise removal. The second feature is to enhance and improves image quality. The method is comprised of four steps. The first step is used for the Cancer imaging Archive (CIA) dataset. The second step is Image noise removal by using a weighted mean histogram. The third step is segmentation using IPCT. Then, feature extraction for lung. Finally, using DITNN for the classification of lung cancer could lead to cancer prediction. The proposed method can detect 98.42% accuracy of lung cancer.

G. Jakimovski et al. [41] stated that many can be saved if lung cancer is diagnosed in the early stages. To identify and classify pulmonary nodules from CT images, they proposed a double convolutional Deep Neural Network (CDNN) and a regular CDNN. A type of novelty has been used in this method. The first novelty is used the K-means algorithm to the image's pre-classification. Next, for cancer research is used the additional convolution layer with edge sharpening filters. The archive of the University of South Carolina and the Laboratory of Neuro Imaging (LONI) datasets are used in this method. The experimental results show that this CDNN has an accuracy of 0.909 and a regular CDNN accuracy of 0.872. Doctors take advantage of this technique to detect and treat lung cancer in early stages.

An automated detection system includes candidate detection and false positive (FP). J. Wang et al. [42] presented a new strategy for the rapid detection of candidates from chest volume CT scans, with minimized false negatives (FNs) and false positives (FPs). They presented a new CAD system for nodule detection constructed from three unified neural networks Including a Feature Extraction Network (FEN), a Region Proposal Network (RPN), and a Region Classification Network (RCN). The TIANCHI AI, LUNA-16, LIDC-IDRI, INDEPENDENT dataset such as Siemens company and General electric company datasets are used in this method. The evaluation of the experiments shows that the presented system is an effective lung cancer screening tool in clinical applications.

C. Wang et al. [43] stated that the most common disease is lung cancer that's why it is the importance of diagnosed in the early stages. They purposed "inception-v3 transfer learning model" for categorizing pulmonary pictures. They amplified the pulmonary imaging data, then applied exact Inception-v3 model based on transfer learning to automatically extract features. The important part of pulmonary image classification is data augmentation. This method is comprised of four steps. The first step is data preprocessing. The second step is the image feature extraction. Next, training the classifier considered features extraction. Then, the model is tested after trained. The experimental results show that this method has a sensitivity of 95.41% of the standard public digital image database JSRT [44]. Pulmonary image classification performance is better than other methods.

In 2018, H. Jiang et al. [45] presented an effective scheme for the detection of pulmonary nodules that used multi-group fragments that were cut out of lung images, augmented by the "Frangi filter". This CADe scheme is comprised of two classes of images and "four-channel convolution neural networks (CNN)". The experimental results show that this method has a sensitivity of 80.06% with 4.7 false positives per scan and a sensitivity of 94% with 15.1 false positives per scan of lung cancer. The LIDC-IDRI dataset is used in this method. It was shown that "multi-group patch-based learning system" is effective on improving the detection process of lung nodes as well as significantly reducing false positives among large amounts of visual data.

Y. Liu et al. [46] proposed a dense convolutional binary-tree network (DenseBTNet). The DenseBTNet is comprised of two attractive advantages. First, the DenseBTNet not only maintains, the DenseNet dense mechanism for extracting the properties of lung nodes at different levels but also strengthens this mechanism to dense blocks level and enhances those features that are multi-scaled. Secondly, parameter efficiency of the "DenseBTNet'' is high and low in parameter scale. The empirical results reveal that "DenseBTNet" greatly increases function of DenseNet and leads to higher accuracy in classifying lung nodes compared to advanced approaches. The LIDC-IDRI dataset is used in this method.

L. Li et al. [47] stated that the main cause of mortality in lung cancer is how it is diagnosed in the early stages. They presented deep learning-based computer-aided diagnosis (DL-CAD) system for detecting and characterizing pulmonary nodules. The DL-CAD system is designed to detect $\geq$ 3 mm nodules and predicts the possibility of malignancy for each node detected. The LIDC-IDRI and the National Cancer Institute NLST datasets are used in this method. The evaluation of the experiments shows that this presented model can detect 86.2 % sensitivity of lung cancer.

An automatic diagnosis of pulmonary nodules using CT scan improves the efficiency of lung cancer diagnosis and false-positive reduction plays an important role in diagnosis. For decreasing false-positive nodules of candidate nodules H. Jin et al. [48] proposed a deep 3D residual convolution neural network (CNN). For extracting multi-level contextual information of CT data, they designed a spatial pooling and cropping (SPC) layer. The SPC layer is more convenient and efficient for this detection. They designed a 27-layer network consisting of the remaining three groups. Each remained group consists of the remaining four units. Each remained unit is comprised of two convolution layers with a core size of $5 \times 5 \times 3$. The LUNA-16 dataset is used in this method. The experimental results show that this method has a sensitivity of 98.3% detecting lung cancer.

In 2017, the main cause of disease and death in lung cancer is how it is diagnosed in the early stages. Accurate and automatic classification of lung nodules is of great importance.
A. Teramoto et al. [14] presented "Deep Convolutional Neural Network (DCNN)" to automatically classify lung cancer. The DCNN is comprised of different layers such as three "convolutional" layers, two "fully connected layers", and three "pooling layers". The seventy-six (76) cases of cancer cells dataset is used in this method. Results display that about 71 percent of the images are classified faultlessly.

An automated detection system includes candidate screening and false positive (FP). For computer-aided detection of pulmonary nodules from volumetric CT scans, Q. Dou et al. [49] proposed 3D convolutional neural networks (CNNs). The 3D CNN is comprised of different layers such as 3D convolutional layers, 3D max-pooling Layer, fully-connected layers, and softmax layer. Each layer is comprised of different channels and each channel encodes a unique pattern. The LUNA16 dataset is used in this method.

The experiments performed on large-scale benchmark datasets demonstrate the effectiveness of 3D CNNs as well as the strategy of integrating multi-level textual information to improve detection accuracy. The proposed model can detect 94.4% sensitivities of lung cancer.

## 4. Analysis of the reviewed of the lung cancer based on deep learning

Three different analyses were performed to evaluate lung cancer based on deep learning:

    The first analysis of classification is based on a literature review (Figure 8).
    The second analysis of architecture, data set, simulation, accuracy and sensitivities are based on literature review (Table 3).
    The third analysis is about the advantages and disadvantages based on literature review (Table 4).

## 4.1. Summary of taxonomy based on literature review

According to the taxonomy presented in the introduction (Figure 2). After reviewing the articles, we explained the relationship of each of these articles with the classification presented in Figure 8.

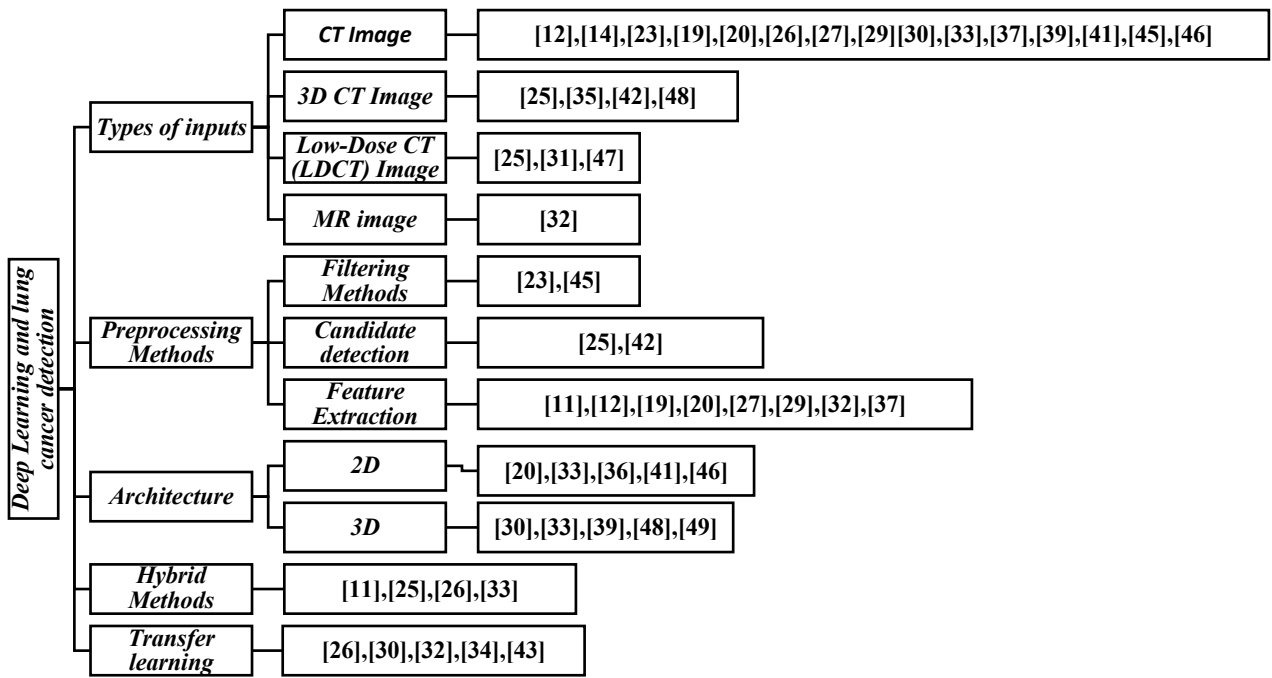

Fig. 8. The Taxonomy of Lung Cancer Detection based on Reviewed Papers

The reviewed articles are classified according to the type of inputs, processing methods, architecture, hybrid methods and transfer learning shown in Figure 8.

## 4.2. Summary of primary studies based on Literature Review

We thoroughly review and analyze early lung cancer studies based on deep learning. Our observations are summarized in Table 3. Table 3 analysis includes architecture, data sets, simulations, accuracy, and sensitivity.

Table 3. An Overview of Existing Primary Studies

| # | Ref | Year | Architecture | Datasets | Evaluation criteria | ACC/ Sen | Simulations |
|---|---|---|---|---|---|---|---|
| 1 | [19] | 2021 | Genotype-guided radiomics method (GGR) | Public radiogenomics dataset of NSCLC | Accuracy | 83.28%/- | PYTHON-Keras |
| 2 | [20] | 2021 | An MLP for the final classification of the EGFR mutation status. | LIDC-IDRI, NSCLC-Radiogenomics | AUC | -/- | - |
| 3 | [12] | 2020 | Adaptive Hierarchical Heuristic Mathematical Model (AHHMM) | http://diagnijmegen.nl/index.php/Lung_Cancer | recognition rate, misclassification ratio, precision, sensitivity, and accuracy | 96.67% / - | - |
| 4 | [22] | 2020 | Low-Dose CT Scans | LIDC-IDRI, LUNA-16, Kaggle | Sensitivity PR AUC true probability ROC AUC | 96.5% / - | - |
| 5 | [23] | 2020 | A Multi-Scene Deep Learning Framework (MSDLF) by the vesselness filter | LIDC-IDRI | Sensitivity ROC Efficiency | - / - | - |
| 6 | [24] | 2020 | A 3D Deep Convolutional Neural Network (3DDCNN) | ANODE09, LUNA-16, LIDC-IDRI, SHANGHAI Hospital | Accuracy Sensitivity Specificity AUROC false positives per scan | 98.5% / - | Intel Extended Caffe |
| 7 | [25] | 2020 | Knowledge-based Analysis of Mortality Prediction Network (KAMP-Net) | National Lung Screening Trial (NLST) | ROC AUC cross-entropy | - / - | PYTHON-PyTorch |
| 8 | [26] | 2020 | Deep learning + DenseNet + Adaboost | Shandong Provincial Hospital | Accuracy Sensitivity precision | 89.85% / - | - |
| 9 | [27] | 2020 | MTMR-Net | https://github.com/CaptainWilliam/MTMR-NET | Accuracy Sensitivity Specificity ROC AUC | 93.5% / - | - |
| 10 | [29] | 2020 | STM-Net | Zhongshan Hospital Fudan University (ZSDB) | Accuracy Sensitivity Specificity ROC AUC | 95.455% / - | PYTHON – Tensorflow |
| 11 | [30] | 2020 | Mask Region-Convolutional Neural Network (Mask R-CNN) | LUNA-16, Ali TianChi challenge | Sensitivity Specificity F-score false positives per scan | - / 88.7% | - |
| 12 | [31] | 2019 | *A multi-view knowledge-based collaborative (MV-KBC)* | *LIDC-IDRI* | *Accuracy sensitivity specificity precision AUC* | 91.60% / - | *Keras-Tensorflow* |
| 13 | [32] | 2019 | *Deep learning for thoracic MR images* | *First Affiliated Hospital of Guangzhou Medical University* | *ROC Sensitivity false positives per scan (FP/scan)* | - / 85.2% | - |
| 14 | [33] | 2019 | *A hybrid segmentation network (HSN)* | *Hospital Affiliated to Shandong University* | *Dice loss Accuracy Sensitivity Precision Dice score* | 0.909 / 0.872 | - |
| 15 | [34] | 2019 | *Multiple view sampling-based multi-section CNN* | *LIDC-IDRI* | *Accuracy sensitivity specificity AUC* | 93.18% / - | *Keras-Tensorflow* |
| 16 | [35] | 2019 | *Nodule-plus Region-based CNN + Deep Self-paced Active Learning (DSAL)* | *LIDC-IDRI* | *Dice loss TP Dice* | - / - | *PYTHON PyTorch* |
| 17 | [36] | 2019 | *Deep reinforcement learning models* | *medical big data* | - | - / - | - |
| 18 | [37] | 2019 | *Optimal Deep Neural Network (ODNN) + Modified Gravitational Search Algorithm (MGSA)* | *http://www.via.cornell.edu* | *Accuracy sensitivity specificity PPV NPV* | 94.56% / 96.2% | *MATLAB* |
| 19 | [39] | 2019 | *Three-dimensional convolutional neural network (CNN)* | *LUNA-16, Kaggle, Guangdong Provincial People's Hospital* | *Accuracy sensitivity specificity* | - / 84.4% | *PYTHON PyTorch* |
| 20 | [13] | 2019 | *Artificial Neural Network (ANN)* | *The lung cancer dataset, named survey lung cancer [40]* | *Accuracy* | 96.67% / - | - |
| 21 | [11] | 2019 | *Improved profuse clustering technique (IPCT) + Deep Learning with Instantaneously Trained Neural Networks (DITNN)* | *Cancer Imaging Archive (CIA)* | *Accuracy Specificity Precision F1 score Error rate Efficiency* | 98.42% / - | *MATLAB* |
| 22 | [41] | 2019 | *Convolutional Deep Neural Network (CDNN) + A regular CDNN* | *Data archive of the university of South Carolina and the Laboratory of Neuro Imaging (LONI)* | *Accuracy Specificity Precision ROC Positive Predictive* | 0.909 and 0.872/ - | *Python Keras* |
| 23 | [42] | 2019 | *A Feature Extraction Network (FEN) + A Region Proposal Network (RPN) + A Region Classification Network (RCN)* | *TIANCHI AI, LUNA-16, LIDC-IDRI, INDEPENDENT datasets that include: 1) Siemens company 2)* | *Free-Response Receiver Operating Characteristic (FROC) Sensitivity* | - / - | *Google Tensorflow* |

| #  | Ref  | Year | Method | Dataset | Metrics | Results | Tool |
|----|------|------|--------|---------|---------|---------|------|
|    |      |      |        | General electric company |  |  |  |
| 24 | [43] | 2019 | Inception-v3 transfer learning | the standard public digital image database JSRT [38] | sensitivity specificity | - / 95.41% | - |
| 25 | [45] | 2018 | Multi-group fragments cut from lung images, augmented by the Frangi filter | LIDC-IDRI | AUC-ROC Sensitivity FP/scan | - / 80.06% and 94% | Matlab |
| 26 | [46] | 2018 | A dense convolutional binary-tree network (DenseBTNet) | LIDC-IDRI | Accuracy AUC | - / - | - |
| 27 | [47] | 2018 | Deep learning-based computer-aided diagnosis (DL-CAD) | LIDC-IDRI and the National Cancer Institute NLST | True positive False positive False negative True negative Sensitivity Specificity | -/ 86.2% | SPSS |
| 28 | [48] | 2018 | A deep 3D residual CNN (convolution neural network) + SPC layer | LUNA-16 | FROC | - / 98.3% | Keras |
| 29 | [14] | 2017 | Deep Convolutional Neural Network (DCNN) | Seventy-six (76) cases of cancer cells | accuracy | 71% / - | Caffe |
| 30 | [49] | 2017 | A 3D CNNs | The LUNA16 Challenge held in conjunction with ISBI 2016 | false positives reduction | - / 94.4% | Python Theano |

In this study, all data sets and simulations that achieve the defined objectives were identified. Check out the results section for more details on data sets and simulations.

### 4.3. Summary of Advantages and Disadvantages based on Literature Review

After reviewing the articles, we stated the main idea, advantages and disadvantages of each article in Table 4.

Table 4. The Comparison of Articles Related to the Lung Cancer based on Deep Learning

| # | Ref | Main Idea | Advantages | Disadvantages |
|---|-----|-----------|------------|---------------|
| 1 | [19] 2021 | NSCLC prediction | 1) High accuracy 2) Low cost | Not mentioned in the relevant article |
| 2 | [20] 2021 | The final classification of the EGFR mutation status. | 1) Obtaining the best prediction ability 2) An innovative approach for gene mutations status assessment | Obtaining a more complete characterization of lung cancer-related genes |
| 3 | [12] 2020 | Build an intelligent and automated cancer prediction system | 1)Automated detection lung cancer 2) Detects the presence of lung cancer 3) likelihood distribution method improved the images quality. | Not mentioned in the relevant article |
| 4 | [22] 2020 | a system called Low-Dose CT Scans for the relationship between algorithmic solutions and physicians | 1) directly analyzes the CT scan and Provides calibrated scores 2) it is developed and configured simultaneously. 3) achieves state-of-the- art performance on lung nodule detection 4) It can reduce cancer mortality by at least 20% annually | The performance is still bounded by the limitations of the datasets used to train our models such as the lack of large nodule annotations |
| 5 | [23] 2020 | The main idea of this analysis is to determine large nodes (> 3 mm). | It is effective in detecting lung nodules to increase accuracy and dramatically reduce false positives in large volumes of image data. | Not mentioned in the relevant article |
| 6 | [24] 2020 | Drawing lung nodules manually by radiologists is a time-consuming and tedious task. That's why they presented with a system to help radiologists | 1)Their system works better than advanced systems. 2)The combination of deep learning and cloud computing is used to accurately detect lung nodules. | 1) The current tested performance metric of 3DDCNN is relatively high 2) The performance was relatively less accurate in detecting micro nodules (diameter is less than 3 mm) |
| 7 | [25] 2020 | Predict the risk of mortality in lung cancer | The data augmentation is effective for improving the performance of CNN | 1)They selected the slices manually 2) The dataset used in their current work is of limited size. 3) The clinical measurements used in this study are manually acquired. 4) The current thresholds and channel arrangement were manually set |
| 8 | [26] 2020 | For solving the low data problem collected by the patient | 1)Extend training data 2) Automated classification model 3) Their model achieved better classification results in CT image classification of malignant tumors | Not mentioned in the relevant article |
| 9 | [27] 2020 | The automatic analysis of pulmonary nodules | Comparing with other advanced methods, they achieved the best accuracy, sensitivity, specificity. | Not mentioned in the relevant article |
| 10 | [29] 2020 | To improve the diagnosis of pulmonary adenocarcinoma | Successfully detection of lung cancer. | Not mentioned in the relevant article |
| 11 | [30] 2020 | *Detection and segmentation for the pulmonary nodule.* | *1) Generalizes well on unseen data 2) Their proposed methods can detect and segment pulmonary nodules more accurately 3) Semi-supervised learning can effectively overcome the problem of small scale of dataset* | *Memory optimization is also a problem to be solved in practical applications.* |
| 12 | [31] 2019 | *Separate benign-malignant lung nodules* | *Their model is more accurate than current state-of-the-art approaches on the LIDC-IDRI dataset.* | *Data annotation.* |
| 13 | [32] 2019 | *Nodule detection for MR images* | *1) This detection scheme can avoid candidate extraction and be less dependent on scale. 2) An FP reduction scheme based on the anatomical characteristics of lung nodule is designed.* | *Some small and low contrast nodules are not detected by Faster R-CNN. 2) Some air artifacts and juxta heart tissues may be falsely detected as nodules.* |
| 14 | [33] 2019 | *Accurate cancer segmentation* | *Their experiments showed that the HSN achieves better performances than other state-of-the-art methods.* | *1) The CT scans were acquired at a single center 2) Scans of healthy people were not included in the dataset 3) Computation and memory requirement.* |
| 15 | [34] 2019 | *Obtaining node structural information from CT scans* | *1) Not require manual spatial annotation 2) lightweight* | *Not mentioned in the relevant article* |
| 16 | [35] 2019 | *For detection of pulmonary nodules visible in 3D CT images* | *1)Reduce training and annotation efficiently 2) Achieve state-of-the-art pulmonary nodule instance segmentation performance* | *Not mentioned in the relevant article* |
| 17 | [36] 2019 | *Use for lung cancer localization and treatment* | *Making the personalized and patient-centered treatment plan* | *Defining an appropriate reward function that is used to update the Q-value for each action.* |
| 18 | [37] 2019 | *For classification Computed Tomography (CT) images of lung* | *1) The manual label time reduction 2) Preventing human error 3) The optimized structure 4) Speedy, simple to operate, non-invasive and cheap* | *Not use high dosage CT lung images* |
| 19 | [39] 2019 | *To identify and classify pulmonary nodules from CT images* | *High sensitivity and high specificity for pulmonary nodule detection and classification* | *1)Relatively few data were currently input 2) The validation data was derived from a multicenter data set, that lead het- erogeneity in image quality* |
| 20 | [13] 2019 | *To determine if lung cancer cells are absent or present in human body* | *solving the problem of classification and prediction* | *Not mentioned in the relevant article* |
| 21 | [11] 2019 | *Improving lung image quality and diagnose lung cancer* | *1) Image noise removal 2) Enhancing and improving image quality* | *Not mentioned in the relevant article* |
| 22 | [41] 2019 | *To identify and classify pulmonary nodules from CT images,* | *1)help in early lung cancer detection and early treatment 2) The highest accuracy* | *Not mentioned in the relevant article* |
| 23 | [42] 2019 | *The rapid detection of candidates* | *1) Detecting nodules and masses with a broad spectrum of appearance, regardless of their types, sizes and locations 2) Simplifying the design of the false positive reduction 3) Candidate detection is very fast* | *1) It may fail to find some nodules 2) Some nodule-like tissues, It may be recognized as nodules* |
| 24 | [43] 2019 | *For categorizing pulmonary pictures* | *1) Pulmonary image classification performance 2) The highest sensitivity and specificity* | *This method is only for lung tasks and is not suitable for other medical images* |
| 25 | [45] 2018 | *An effective scheme for the detection of pulmonary nodules* | *Using 2D image information is a more ideal way to detect the lung nodules.* | *Selecting manually parameters* |
| 26 | [46] 2018 | *Lung nodule diagnostic classification* | *1) DenseBTNet can learn more compact and accurate models than DenseNet 2) DenseBTNet exceeds state-of-the-art approaches in lung nodule malignancy suspiciousness classification.* | *Not mentioned in the relevant article* |
| 27 | [47] 2018 | *For detecting and characterizing pulmonary nodules* | *1) Detecting ≥ 3 mm nodules 2) Predicts the possibility of malignancy for each node detected* | *Not mentioned in the relevant article* |
| 28 | [48] 2018 | *For decreasing false-positive nodules of candidate nodules* | *1) Achieve high nodule detection sensitivity 2) Extracting multi-level contextual information* | *The computational complexity* |
| 29 | [14] 2017 | *Automatically classify lung cancer* | *Classification of lung cancer in cytodiagnosis* | *Not mentioned in the relevant article* |
| 30 | [49] 2017 | *Detection of pulmonary nodules* | *Improving detection accuracy* | *Not mentioned in the relevant article* |

## Results

Having analysed and reviewed the mentioned articles, questions raised earlier in the article will be responded to:

**Answer to Question RQ1:**

To collect survey articles, terminology review, survey, mapping, and literature were searched in the article titles. After reading the abstracts, we tagged the articles. Out of 125 available articles, no articles were found as SLR or SMS. The frequency of survey articles expressed in the titles is shown in Figure 9.

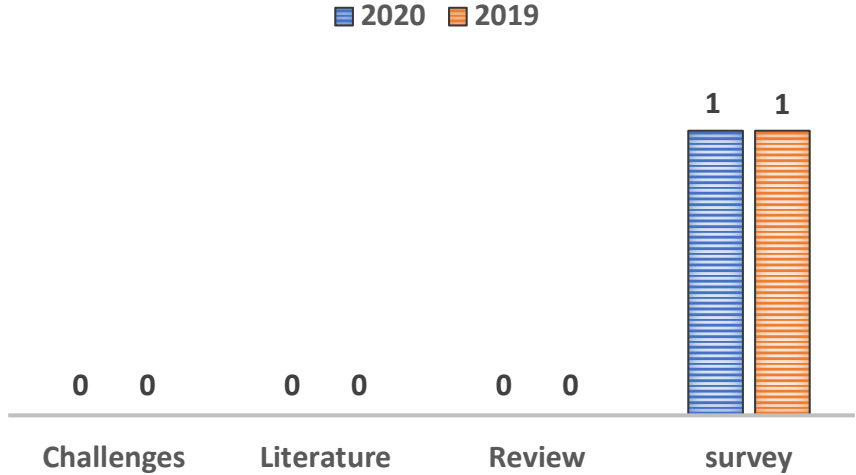

Fig. 9. Type of Article on the Lung Cancer

Survey articles list is shown in Table 5.

Table 5. List of Survey Articles

| # | Authors | Title | Publisher | Year |
|---|---|---|---|---|
| 1 | P. Monkam; S. Qi; H. Ma; W. Gao; Y. Yao; W. Qian | *Detection and Classification of Pulmonary Nodules Using Convolutional Neural Networks: A Survey* [50] | *IEEE journal* | 2019 |
| 2 | R. Mastouria, N. Khlifa, H. Neji, S. H. Zannad | *Deep learning-based CAD schemes for the detection and classification of lung nodules from CT images: A survey* [51] | *IEEE journal* | 2020 |

**Answer to Question RQ2:**

The number of articles on lung cancer in their titles and based on deep learning in the mentioned databases was 125. Figure 10 shows the corresponding percentage in the five selected databases.

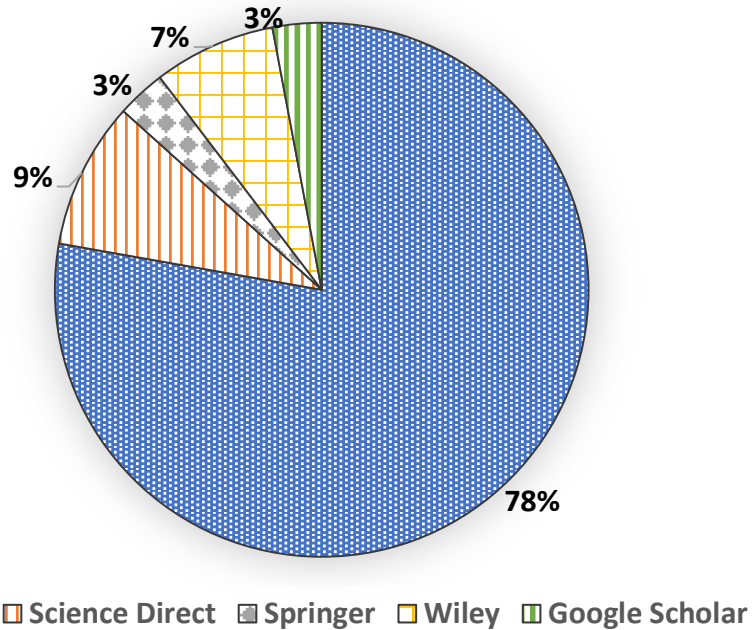

Fig. 10. The Percentage of Articles on Lung Cancer by Five Intended Databases

As shown in Figure 10, the highest percentage of database of our study on lung cancer based on deep learning is IEEE Journal and Table 6 shows their number of five intended databases.

Table 6. The Number of Articles by Five Intended Databases.

| # | Data Base | Number |
|---|---|---|
| 1 | IEEE | 105 |
| 2 | Science Direct | 12 |
| 3 | Springer | 4 |
| 4 | Wiley | 10 |
| 5 | Google Scholar | 4 |

**Answer to Question RQ3:**
To answer this question, we reviewed the simulation that submitted the 32 articles used in this study. Table 7 lists the different simulations.

Table 7. List of Various Simulation in the Articles

| # | Ref | Simulations |
|---|---|---|
| 1 | A. Masood et al. [24] | Intel Extended - Caffe |
| 2 | H. Guo et al. [25] | PYTHON - PyTorch |
| 3 | J. Zheng et al. [29] | PYTHON –Tensorflow |
| 4 | Y. Xie et al. [31] | Keras -Tensorflow |
| 5 | P. Sahu et al. [34] | Keras -Tensorflow |
| 6 | W. Wang et al. [35] | PYTHON - PyTorch |
| 7 | Lakshmanaprabu S.K et al. [37] | MATLAB |
| 8 | C. Zhang et al. [39] | PYTHON - PyTorch |
| 9 | P. M. Shakeel et al. [11] | MATLAB |
| 10 | G. Jakimovski et al. [41] | PYTHON - Keras |
| 11 | J. Wang et al. [42] | Google Tensorflow |
| 12 | H. Jiang et al. [45] | MATLAB |
| 13 | L. Li et al. [47] | SPSS |
| 14 | H. Jin et al. [48] | Keras |
| 15 | A. Teramoto et al. [14] | Caffe |
| 16 | Q. Dou et al. [49] | PYTHON - Theano |
| 17 | P. Aonpong et al. [19] | PYTHON - Keras |

Python is commonly used in deep learning articles on lung cancer implemented using PyTorch's open source library in Python [52]. Figure 11 shows a comparison between Python, MATLAB, Caffe, PyTorch, Keras, Tensorflow, Theano, SPSS, and Intel Extended that have been used in deep learning lung cancer articles.

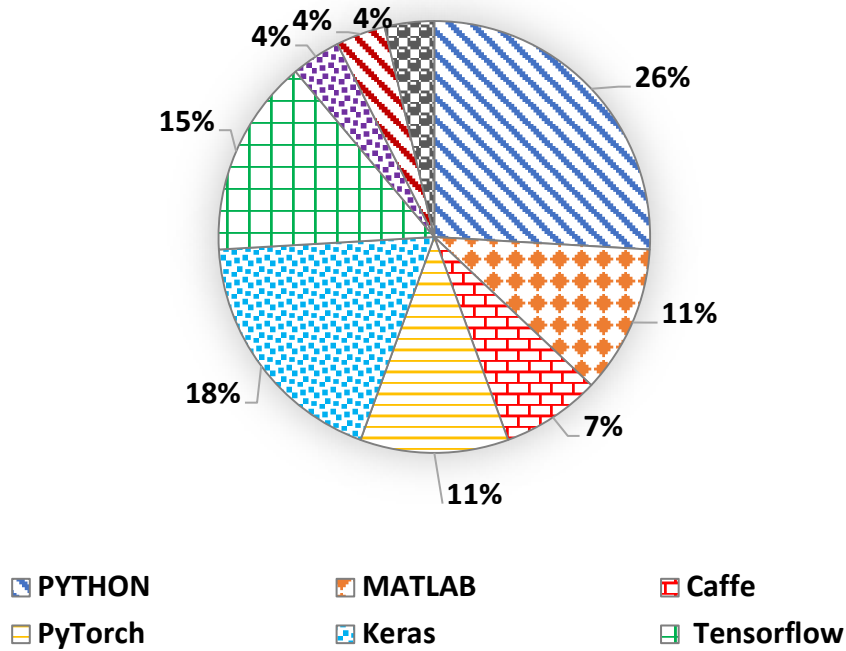

Fig. 11. The Percentage of Simulations on Lung Cancer based on Deep Learning.

As shown in Figure 11, the highest percentage of simulations our study on lung cancer based on deep learning is Python. This indicates that python is useful simulation.

**Answer to Question RQ4:**
To respond to this question, we reviewed the dataset that submitted 32 articles used in this study. As shown in Table 3, we have thoroughly reviewed the dataset in the articles. The best data set is shown in Figure 12 as shown in Table 3.

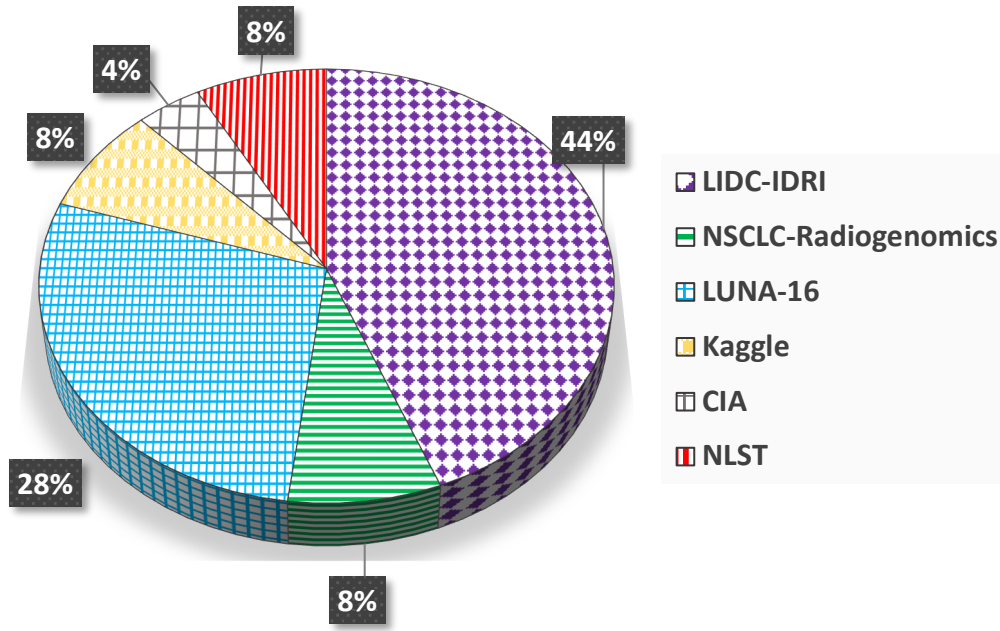

Fig. 12. The Percentage of Information on the Datasets

According to Figure 12, the best data set studied is plotted in 32 articles. We will now briefly describe each of these datasets.

LIDC-IDRI [53]: LIDC-IDRI: LIDC-IDRI is a collection of lung cancer screening data that includes chest CT scans for a total of 1,010 patients.

NSCLC-Radiogenomics [54]: NSCLC-Radiogenomics is a collection of publicly available data with CT images for a group of patients with Non-Small Cell Lung Cancer (NSCLC). Also, NSCLC-Radiogenomics is the only general data set consisting of paired information about the status of gene mutations associated with lung cancer.

LUNA-16 [55]: LUNA16 (Lung Node Analysis) is a data set for lung segmentation consisting of 1,186 lung nodes detailed in 888 CT scans.

Kaggle [56]: Kaggle allows users to find and publish datasets.

CIA [57]: The Cancer Imaging Archive (TCIA) is a free access database of medical images for cancer research. The site is funded by the National Cancer Institute Cancer Imaging Program, and is administered by University of Arkansas for Medical Sciences.

NLST [58]: The National Lung Screening Trial (NLST) is lung cancer dataset that compares low-dose Computed Tomography (CT) with chest radiography.

**Answer to Question RQ5:**
To answer this question, we surveyed the universities that submitted 32 articles used in this study. Table 8 shows the superiority of the universities that worked in this field.

Table 8. Universities that worked on the Lung Cancer based on Deep Learning

| University Name | Country |
|---|---|
| Deparment of Electrical and Computer Engineering, Stevens Institute of Technology, Hoboken, NJ | USA |
| Department of Computer Science, Stony Brook University, Stony Brook, NY | USA |
| Department of Computing, Imperial College London, London | U.K. |
| School of Information Technologies, The University of Sydney, Sydney, NSW | Australia |
| School of Software Engineering, Tongji University, Shanghai | China |
| Department of Biomedical Engineering, Shandong University, Jinan | China |
| Department of Computer Science and Engineering, The Chinese University of Hong Kong, Hong Kong; | China |

Figure 13 shows the country's share of published articles on lung cancer based on in-depth learning.

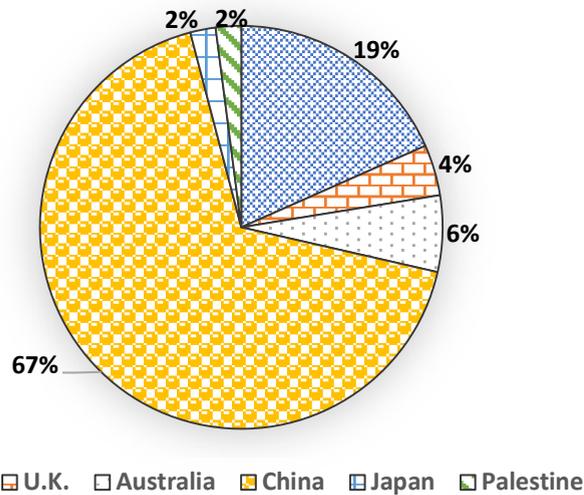

Fig. 13. Percentage of Articles on Lung Cancer based on Deep Learning based on Country Share

**Answer to Question RQ6:**
Deep learning is more popular than traditional methods of image classification because deep learning algorithms have recently been considered as a promising method in medicine such as DCNN [14] adaboost [26] MTMR-Net [27] STM-Net [29] ray-casting volume rendering [30] MV-KBC[31] and ODNN [37], inception-v3 transfer learning [43]. Soon, deep learning must go beyond machine learning in image classification, object recognition, and feature extraction. According to the 32 articles reviewed, table 9 shows the deep learning algorithms used to classify an image.

Table 9. Types of Classification Algorithms in Deep Learning

| # | Ref | Algorithms |
|---|---|---|
| 1 | A. Teramoto et al. [14] | DCNN |
| 2 | S. Pang et al. [26] | adaboost |
| 3 | L. Liu et al. [27] | MTMR-Net |
| 4 | J. Zheng et al. [29] | STM-Net |
| 5 | L. Cai et al. [30] | ray-casting volume rendering |
| 6 | Y. Xie et al. [31] | MV-KBC |
| 7 | Lakshmanaprabu S.K et al. [37] | ODNN |
| 8 | C. Wang et al. [43] | inception-v3 transfer learning |

**Answer to Question RQ7:**

As in Question 6, we examined the different algorithms used in 32 articles, we want to examine the accuracy or sensitivity obtained from the algorithms used in the architectures of these 32 articles. Table 10 shows the accuracy or sensitivity used in the architectures of these 32 articles.

Table 10. The Percentage of Accuracy or Sensitivity Classification Algorithms in Deep Learning.

| # | Architectures | Accuracy / sensitivity |
|---|---|---|
| 1 | Deep Convolutional Neural Network (DCNN) | 71% / - |
| 2 | Deep learning + DenseNet + Adaboost | 89.85% / - |
| 3 | MTMR-Net | 93.5% / - |
| 4 | STM-Net | 94.455% / - |
| 5 | Mask Region-Convolutional Neural Network (Mask R-CNN) | - / 88.7% |
| 6 | A multi-view knowledge-based collaborative (MV-KBC) | 91.60% / - |
| 7 | Optimal Deep Neural Network (ODNN) + Modified Gravitational Search Algorithm (MGSA) | 94.56% / 96.2% |
| 8 | inception-v3 transfer learning | - / 95.41% |

Figure 14 shows the different percentages of accuracy or sensitivity of algorithms used in the architectures.

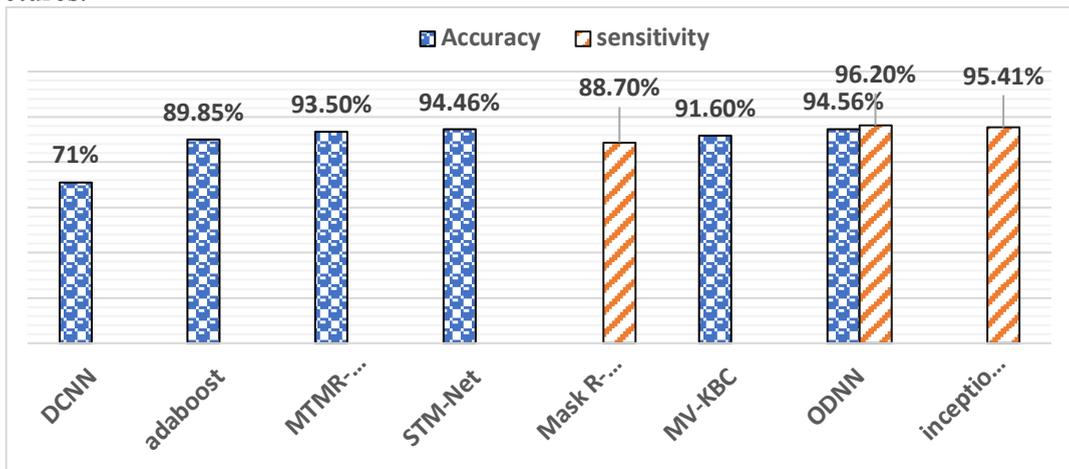

Fig. 14. The Percentage of Algorithms of Accuracy or Sensitivity in these Architectures.

## 5. Conclusion

In the present paper, the results of the SMS with SLR on lung cancer based on deep learning are presented. Thirty-two articles were reviewed and various methods for diagnosing lung cancer were presented with deep learning, which resulted in different levels of accuracy and sensitivity. This article examines the universities and countries that have contributed to the publication of related articles. Most of the publications in this field are attributed to the IEEE. China, the United States, Australia, the United Kingdom, Japan, and Palestine had the largest number of articles, respectively.

Current article contains limitations since searches were done in the title of the articles only and by expanding this search, other articles can be accessed.

Focus of this study has been on English publications only and has not done any search on non-English papers.

## Acknowledgement

The authors would like to thank you Dr. Armin Chitizade for reviewing the manuscript and Dr. Ghasem Naghib for general guidance through this research.

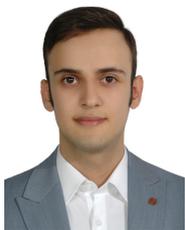

**Hesamoddin Hosseini** is M.Sc. of Student in Artificial Intelligence and Robotics from Ferdowsi University of Mashhad, Mashhad, Iran. He received his Bachelor of Computer Software Engineering with an Honors Class 1 in 2021 from Islamic Azad University (IAU) Mashhad. He obtained the top student award from the School of Engineering at 2021 IAUM Engineering. He has been included in the Engineering Dean's Honors lists for years 2019, 2020 and 2021. His research interests include the Medical Image processing, Biomedical Image Analysis, Machine Learning and Deep Learning.

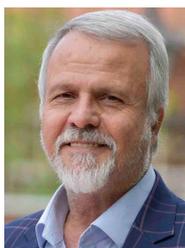

**Reza Monsefi** born in Ahwaz year 1956 in south of Iran and has received his Honors Degree in Electrical and Electronic Engineering from Manchester University, 1978, Manchester, U.K., M.Sc. in Control Engineering, Salford University,1981, Manchester, U.K. and Ph.D. in Data Communication and Supervisory Control, 1983, Salford University, Manchester, U.K. He is a fellow member of IEE, and at present a senior lecturer, professor and works with Ferdowsi University of Mashhad, Mashhad, Iran. Computer Networks, Wireless Sensor Networks and Machine Learning and Soft Computing are among his professional interests

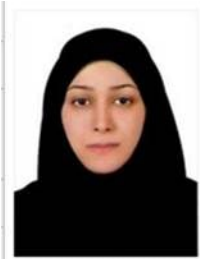

**Shabnam Shadroo** received her B.S. degree in Hardware Computer Engineering from the University of Sajad, Mashhad, in 2005, the MS in Artificial Intelligence Computer Engineering from the IAU, Mashhad, in 2007 and the Ph.D. degree in Software Computer Engineering from SRBIAU University, Tehran, in 2021. She joined the Department of software computer Engineering, Azad University of Mashhad, as a Lecturer, in 2006 and was promoted to a Faculty member in 2009. Her current research interests include machine vision, deep learning, Big data, and the internet of things.